\documentclass{article}
\usepackage[superscript,biblabel]{cite}

\begin{document}

\title{Grover's Algorithm in Natural Settings}
\author{Apoorva D. Patel$^*$\\
        Centre for High Energy Physics, Indian Institute of Science,\\
        Bangalore 560012, India\\
        $^*$adpatel@iisc.ac.in\\}

\date{}
\maketitle

\begin{abstract}
The execution of Grover's quantum search algorithm needs rather limited
resources without much fine tuning.
Consequently, the algorithm can be implemented in a variety of physical
set-ups, which involve wave dynamics but may not need other quantum features.
Several of these set-ups are described, pointing out that some of them occur
quite naturally.
In particular, it is entirely possible that the algorithm played a key role
in the selection of the universal structure of genetic languages.
\end{abstract}

\section{Grover's Algorithm}

Lov Grover discovered a marvellous algorithm for unstructured search in
the context of quantum computation \cite{grover1996}.
Formally, the problem is to find a target item with specific properties
in an unsorted database using a set of binary queries.
The algorithm starts with a uniform superposition state, and alternately
applies two reflection operators for a number of iterations, until the
target state is reached.
One of the reflection operators is the response to the binary query,
the other is the reflection across the uniform state, and the number
of iterations needed to reach the target state is $O(\sqrt{N})$ for a
database of size $N$.
Any Boolean algorithm would require $O(N)$ binary queries to accomplish
the same task starting from an unbiased state, so this is a square-root
improvement in the computational efficiency.
Furthermore, the algorithmic evolution is at a constant rate along the
geodesic from the initial state to the final state, taking place in the
two-dimensional subspace (of the total $N$-dimensional space) formed by
the uniform state and the target state.
That makes it the optimal solution to the problem \cite{zalka1999}.

The simplicity of the algorithm makes it implementable in a variety of
physical settings, and a number of its variations and applications have
been explored over the years \cite{qsreview}.
The key feature of the algorithm is wave dynamics that allows
superposition; other quantum features can be easily skipped.
Once coherent wave modes are available, the algorithm needs nothing
more than suitable reflection operations.
Figure 1 illustrates how the algorithm works in the simplest case,
unambiguously identifying one out of four items in the database using
a single binary oracle call.
In contrast, a single binary oracle call in a Boolean setting would
only identify one out of two items in the database.
Note that Grover's algorithm is referred to as a search algorithm due to
the quantum interpretation of $|{\rm amplitude}|^2$ as probability.

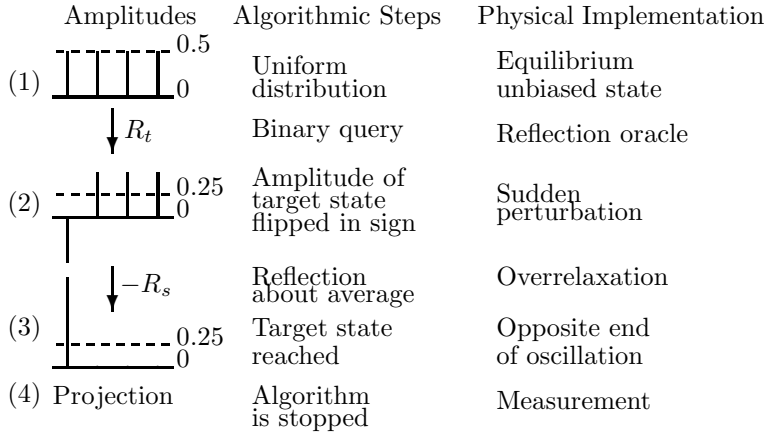
\begin{figure}[t]
\setlength{\unitlength}{0.5mm}
\hspace{5mm}
\begin{picture}(150,120)
  \thicklines
\put(15,105){\makebox(0,0)[bl]{Amplitudes}}
\put(60,105){\makebox(0,0)[bl]{Algorithmic Steps}}
\put(125,105){\makebox(0,0)[bl]{Physical Implementation}}
  \put( 0,87){\makebox(0,0)[bl]{(1)}}
  \put(12,87){\line(1,0){32}}
\put(13,99){\line(1,0){2}} \put(17,99){\line(1,0){2}} \put(21,99){\line(1,0){2}}
\put(25,99){\line(1,0){2}} \put(29,99){\line(1,0){2}} \put(33,99){\line(1,0){2}}
\put(37,99){\line(1,0){2}} \put(41,99){\line(1,0){2}}
  \put(45,87){\makebox(0,0)[bl]{0}} \put(45,99){\makebox(0,0)[bl]{0.5}}
  \put(16,87){\line(0,1){12}} \put(24,87){\line(0,1){12}}
  \put(32,87){\line(0,1){12}} \put(40,87){\line(0,1){12}}
  \put(65,93){\makebox(0,0)[bl]{Uniform}}
  \put(65,87){\makebox(0,0)[bl]{distribution}}
  \put(130,93){\makebox(0,0)[bl]{Equilibrium}}
  \put(130,87){\makebox(0,0)[bl]{unbiased state}}
  \put(28,84){\vector(0,-1){12}}
  \put(31,75){\makebox(0,0)[bl]{$R_t$}}
  \put(65,75){\makebox(0,0)[bl]{Binary query}}
  \put(130,75){\makebox(0,0)[bl]{Reflection oracle}}
  \put( 0,55){\makebox(0,0)[bl]{(2)}}
  \put(12,55){\line(1,0){32}}
\put(13,61){\line(1,0){2}} \put(17,61){\line(1,0){2}} \put(21,61){\line(1,0){2}}
\put(25,61){\line(1,0){2}} \put(29,61){\line(1,0){2}} \put(33,61){\line(1,0){2}}
\put(37,61){\line(1,0){2}} \put(41,61){\line(1,0){2}}
  \put(45,55){\makebox(0,0)[bl]{0}} \put(45,61){\makebox(0,0)[bl]{0.25}}
  \put(16,55){\line(0,-1){12}} \put(24,55){\line(0,1){12}}
  \put(32,55){\line(0,1){12}} \put(40,55){\line(0,1){12}}
  \put(65,62){\makebox(0,0)[bl]{Amplitude of}}
  \put(65,56){\makebox(0,0)[bl]{target state}}
  \put(65,50){\makebox(0,0)[bl]{flipped in sign}}
  \put(130,59){\makebox(0,0)[bl]{Sudden}}
  \put(130,53){\makebox(0,0)[bl]{perturbation}}
  \put(28,42){\vector(0,-1){12}}
  \put(30,34){\makebox(0,0)[bl]{$-R_s$}}
  \put(65,37){\makebox(0,0)[bl]{Reflection}}
  \put(65,31){\makebox(0,0)[bl]{about average}}
  \put(130,37){\makebox(0,0)[bl]{Overrelaxation}}
  \put( 0,22){\makebox(0,0)[bl]{(3)}}
  \put(12,15){\line(1,0){32}}
\put(13,21){\line(1,0){2}} \put(17,21){\line(1,0){2}} \put(21,21){\line(1,0){2}}
\put(25,21){\line(1,0){2}} \put(29,21){\line(1,0){2}} \put(33,21){\line(1,0){2}}
\put(37,21){\line(1,0){2}} \put(41,21){\line(1,0){2}}
  \put(45,15){\makebox(0,0)[bl]{0}} \put(45,21){\makebox(0,0)[bl]{0.25}}
  \put(16,15){\line(0,1){24}}
  \put(24,15){\circle*{1}} \put(32,15){\circle*{1}} \put(40,15){\circle*{1}}
  \put(65,22){\makebox(0,0)[bl]{Target state}}
  \put(65,16){\makebox(0,0)[bl]{reached}}
  \put(130,22){\makebox(0,0)[bl]{Opposite end}}
  \put(130,16){\makebox(0,0)[bl]{of oscillation}}
  \put( 0,4){\makebox(0,0)[bl]{(4)}}
  \put(12,4){\makebox(0,0)[bl]{Projection}}
  \put(65,4){\makebox(0,0)[bl]{Algorithm}}
  \put(65,-2){\makebox(0,0)[bl]{is stopped}}
  \put(130,4){\makebox(0,0)[bl]{Measurement}}
\end{picture}
\caption{The steps of Grover's search algorithm for the simplest case of
four items in the database, when the first item is marked by the oracle.
The left column depicts the amplitudes of the four states that evolve
coherently, with the dashed lines showing their average values.
The middle column describes the algorithmic steps,
and the right column mentions their physical implementation.
\label{onefromfour}}
\end{figure}

The algorithm is robust to several types of modifications.
One possibility is to replace the initial uniform state by a generic state.
Then the only change required in the algorithm is to replace the reflection
across the uniform state by the reflection across the specified generic state.
The number of binary queries needed in the algorithm is then of the order
of the reciprocal of the overlap between the target state and the initial
generic state.
This form of the algorithm is referred to as ``amplitude amplification",
which can be used as a subroutine to enhance small success probability
of another algorithm that produces the generic state as an output
\cite{brassard2002}.

Another possibility is to consider the situation where the database is
spread out in space over distinct locations.
In this spatial search problem \cite{aaronson2005}, the items are
represented as vertices of a graph, and there is a restriction that
while searching for the target item one can proceed from one vertex
to the next one only along the edges of the graph.
Grover's algorithm corresponds to the maximally connected graph,
i.e. there is an edge between any two vertices of the graph.
When the graph connectivity is reduced, the reflection in the uniform
state operation has to be replaced by a quantum walk proceeding along
neighbouring vertices.
That decreases the efficiency of the search process.
Still, the square-root improvement in computational efficiency survives
for graphs of effective dimensionality larger than two, while two is the
critical dimension and the square-root improvement is modified there by
a logarithmic overhead \cite{patel2010}.

\section{Grover's Algorithm as Hamiltonian Evolution}
\label{Groverevol}

Grover constructed his algorithm with the physical intuition about the
evolution of a quantum state, where the potential energy term in the
Hamiltonian attracts the state towards the target state and the kinetic
energy term in the Hamiltonian diffuses the state over the whole database
\cite{grover2001}.
Following the Dirac notation, let the target state be $|t\rangle$.
Then the projection operator $P_t=|t\rangle\langle t|$ represents the
potential energy.
Also, let the uniform superposition state be $|s\rangle$.
Then the projection operator $P_s=|s\rangle\langle s|$ represents the
isotropic kinetic energy.
The reflection operators used in the algorithm are easily expressed
in terms of these projection operators as $R_t=I-2P_t$ and $R_s=I-2P_s$.
Grover's algorithm is then the discrete Trotter formula \cite{evolop},
\begin{equation}
|t\rangle = (-R_s R_t)^Q |s\rangle ~,
\end{equation}
which solves the problem with $Q$ queries.
This structure clarifies the reasons behind the extraordinary properties
of the algorithm:
(a) Reflections are the largest steps that one can take consistent with
unitarity, and that makes the algorithm optimal.
(b) The Trotter formula structure allows changes in the strengths of
potential and kinetic energy terms to be largely compensated by a change
in the number of queries, and that makes the algorithm robust.

The evolution of the quantum state remains confined to the two-dimensional
subspace formed by the states $|t\rangle$ and $|s\rangle$.
Let $\langle s|t\rangle \equiv \cos\theta \in [0,1]$ denote the overlap
between these two states; for the uniform initial state
$\cos\theta=1/\sqrt{N}$.
Then we can express
\begin{equation}
|t\rangle = {1 \choose 0} ,~~
|t_\perp\rangle = {0 \choose 1} ,~~
|s\rangle = {\cos\theta \choose \sin\theta} ,
\end{equation}
in the two-dimensional subspace.
Grover's algorithm iterates the discrete evolution operator,
\begin{eqnarray}
\label{operG}
U_G = -R_s R_t
    = \pmatrix{ 1-2\cos^2\theta        & 2\cos\theta\sin\theta \cr
                -2\cos\theta\sin\theta & 2\sin^2\theta-1       }
    = -\cos(2\theta) I + i\sin(2\theta) \sigma_2 ~,
\end{eqnarray}
which rotates the state by the angle $\pi-2\theta$ in the two-dimensional
subspace.
It corresponds to the effective Hamiltonian evolution,
\begin{equation}
U_G \equiv \exp(-iH_G\tau) ~,~~
H_G\tau = (2\theta-\pi) \sigma_2 ~.
\end{equation}
The number of queries required by the algorithm is therefore,
\begin{equation}
Q = \frac{\theta}{\pi-2\theta} \approx \frac{\pi}{4}\sqrt{N} ~.
\end{equation}
This result can be also expressed as:
\begin{equation}
\label{querysoln}
(2Q+1)\sin^{-1}\frac{1}{\sqrt{N}} = \frac{\pi}{2} ~.
\end{equation}
Furthermore, it is clear that if the algorithm is not stopped after $Q$
queries, it keeps on rotating the state at a constant rate in the
two-dimensional subspace, resulting in an oscillatory behaviour of the
amplitude at the target state.

This analysis also makes the spectral properties of the problem obvious.
The kinetic energy operator in the Hamiltonian, $P_s$, has a single
eigenvalue equal to one with eigenvector $|s\rangle$, and $N-1$ eigenvalues
equal to zero associated with the remaining orthogonal directions.
When the potential energy is included in the Hamiltonian, this spectrum
gets modified.
$H_G\tau$ has two eigenvalues equal to $\pm(2\theta-\pi)$ with
(unnormalised) eigenvectors $(|t\rangle \pm i|t_\perp\rangle) \propto
(e^{\pm i\theta}|t\rangle - |s\rangle)$, and $N-2$ eigenvalues equal to
zero associated with the remaining orthogonal directions.
Thus introduction of the potential creates a bound state in the spectrum
with its amplitude concentrated at the target state.
Grover's algorithm is a scattering process in this framework, which focuses
the initially uniform amplitude at the location of the scatterer.
Note that both attractive and repulsive potentials produce the same effect,
since $e^{+i\pi}=e^{-i\pi}$ in construction of $R_t$.

\section{Localisation in Condensed Matter Systems}

Localisation of electron states, due to disorder in a conducting material,
is a well-established phenomenon in condensed matter systems
\cite{anderson1958,abrahams2010}.
The phenomenon has been analysed in detail in the context of a
metal-insulator transition, to demonstrate its genuine quantum nature
(in contrast to classical diffusion).
The disorder can arise from impurities or defects in the material,
and the resultant scatterings impede transport due to interference
among many electron propagation paths.

A prototype model is provided by the tight-binding Hamiltonian:
\begin{equation}
H = \sum_i E_i c_i^\dagger c_i
  - t \sum_{<i,j>} (c_i^\dagger c_j + c_j^\dagger c_i) ~.
\end{equation}
Here $E_i$ denotes the potential energy for the electron at site $i$,
and $t$ is the hopping parameter for the electron to jump from site $i$
to site $j$.
When all the $E_i$ are the same, the spectrum of this Hamiltonian is a set
of energy bands for the electron, and the system is a conducting metal
when the valence band is partially filled.

When one of the $E_i$ is different than the rest, there is a delta-function
potential, say at $i=0$.
Such an attractive potential produces a bound state, separated from the
continuous energy band and localised at $i=0$, for any strength of the
potential in one space dimension and potential strengths beyond a particular
threshold in higher space dimensions.

When the energy disorder has a nonzero density, e.g. $E_i$ are uniformly
distributed over a finite interval, all states can get localised, turning
the conducting system into an insulator.
That happens for any nonzero disorder in one and two space dimensions,
and for sufficiently large disorder (i.e. the magnitude of variation of
$E_i$) in higher dimensions.

Weak localisation is a precursor to the phenomenon described above,
in which the disorder is limited and the associated localisation increases
the resistivity of the material \cite{altshuler1980,hikami1980,abrahams2010}.
It is understood as enhanced probability for electron paths containing closed
loops, due to constructive interference between contributions that travel
the loops in opposite sense.
On the other hand, the paths corresponding to random diffusive motion suffer
destructive interference.
The heightened tendency for electrons to wander around in loops then
increases the resistivity.
Random walks are much more likely to self-cross in lower dimensions than in
higher dimensions; so weak localisation is found strongly in systems of
one and two space dimensions.

Grover's algorithm can be looked upon as a still weaker version of
localisation.
The target site is the only defect, and reflection from it produces a bound
state around it in the spectrum.
The localisation effect is maximised because the evolution dynamics is
restricted to the lowest possible dimensionality, i.e. the subspace formed
by $|s\rangle$ and $|t\rangle$.
Overall, the scattering does not stop propagation of the initial state;
instead the amplitude to be at the target site goes through periodic ups
and downs as a function of time.
Stopping the algorithm at the right time then results in the state having
a large amplitude at the target site.

The modification of Grover's algorithm to spatial search with multiple
target items would be closer to the transport behaviour of a material with
many defects.
In this case also, it is found that introduction of a potential that causes
reflection from the target sites (with all the reflection phase-shifts
chosen to be the same for simplicity) creates an eigenstate localised around
them, and the scattering amplitude to this eigenstate undergoes periodic
ups and downs as a function of time \cite{abhijith2018}.

\section{Variety of Implementations of Grover's Algorithm}

The ingredients required to implement Grover's algorithm are quite simple;
quintessential quantum features such as complex numbers and entangled
states do not explicitly appear.
It suffices to have coherent wave modes that can be superposed and
phase-shifted, and so the algorithm can be implemented using classical
wave dynamics as well \cite{grover2002,patel2006}.
As illustrated in Figure 1, the required features are:\\
(1) an initial state that is correlated in phase among its wave modes,\\
(2) a reflection oracle that singles out the target state,\\
(3) coherent oscillations of the wave modes about the direction specified
by the initial state, and\\
(4) a threshold trigger that stops the algorithm when the target
state amplitude becomes sufficiently large.

These features can be found in a variety of physical settings.\\
(1) It is known since the time of Huygens, that pendulums suspended on
a single wall automatically synchronise.
The tiny coupling between the oscillators provided by the common support
of the wall suffices for this purpose.
Such synchronisation of oscillations has been observed in nanoscale
systems too \cite{colombano2019}.
The synchronised state is an equilibrium state, and so is well-suited
to be the initial state of the algorithm.\\
(2) Any object with properties distinct from the rest can be looked upon
as an impurity. 
Impurities in a material generically scatter wave modes.
When the impurity is a node for wave propagation, reflection from it
changes the sign of the wave amplitude, as can be easily deduced using the
method of images.\\
(3) When the initial state is an equilibrium state, the perturbation
caused by the target oracle would naturally produce oscillations
about the equilibrium direction.\\
(4) There exist many phenomena and reactions that need a critical
threshold to be crossed. 
They can be rapidly completed by amplitude amplification, with the threshold
crossing becoming an effective measurement that terminates the algorithm.
(In such a situation, the number of iterations in the algorithm is not
decided at the outset; instead the iterations continue until success.)

Furthermore, these features are fairly immune to variations.
As a result, Grover's algorithm has been extended to a multitude of
physical scenarios since its discovery, and still found to do its job.
Several realisations that have been pointed out, in addition to the
localisation phenomenon described in the previous section, are:

\noindent
$\bullet$ A coupled system of classical oscillators, with dynamics far
sturdier against environmental disturbances than the quantum case,
can execute Grover's algorithm \cite{grover2002,patel2006}.
The centre-of-mass mode plays the role of the uniform superposition state,
and the reflection operations are implemented as elastic collisions.
The frequencies have to be chosen to allow resonant amplitude transfer,
and a high school science project has demonstrated the scenario
\cite{patel2007a}.
A point to note is that both the classical wave version and the quantum
version of the algorithm have identical oracle complexity, but the
classical version needs $N$ distinct wave modes while the quantum one
requires $\log_2N$ qubits.
The classical wave version therefore needs more spatial resources for the
algorithm than the quantum one, although the temporal resources are the
same in both cases.
Another property is that $|{\rm amplitude}|^2$ represents energy in the
mechanical setting, in contrast to it representing probability in the
quantum setting.
So the mechanical version of the algorithm provides the optimal method to
focus energy, or while running in reverse, the optimal method to disperse
energy.
(Observe that energy is neither supplied nor extracted during the running
of the algorithm.)
The consequences can be dramatic in processes whose rates are governed by
the Boltzmann factor, $\exp(-E/kT)$, where the energy appears in the
exponent.
Efficient schemes to transfer/redistribute energy have many practical
uses in systems ranging from mechanical to electrical, chemical and
biological ones.
Some possibilities are: focusing of energy can be used as a selective
switch, energy amplification can speed up catalytic processes, defects
and impurities in materials can be detected by wave reflections at
suitably tuned frequencies, and fast dispersal of energy can be used in
shock absorbers.

\noindent
$\bullet$ As mentioned earlier, the Trotter formula structure of Grover's
algorithm suggests that the reflection operations can be replaced by 
other values of phase-shifts or related operations, and the algorithm
will still succeed, albeit with somewhat reduced efficiency.
In spatial search problems, the reflection across the uniform state is
replaced by a quantum walk generated by the discrete Laplacian operator,
and an extra coin degree of freedom controls the choice of movement
directions.
The best algorithms involve relativistic quantum walks, with the coin
becoming the inherent internal degree of freedom \cite{portugal}.
The Klein-Gordon version as well as the Dirac version of the quantum walk
have been studied; the former is easily extended to fractal geometries
as well as to search for multiple targets.
A general framework with the reflection across the uniform state replaced
by any diffusion operator is analysed as well, where the spectral properties
of the diffusion operator decide the advantage provided by the algorithm
\cite{tulsi2012}.

\noindent
$\bullet$ 
The reflection oracle for the target state can be also replaced by a
phase-shift different from $\pi$.
Then the optimal performance is obtained, as expected from the Trotter
formula structure, when this phase-shift matches the phase rotation
provided by the diffusion operator \cite{tulsi2012}.
In spatial search problems, the phase-shift at the target state can also
be created in many ways: scattering from a localised potential,
scattering from an obstacle \cite{guillet2016},
localized change in the effective mass of the propagating mode
\cite{tulsi2008,patel2010a},
closed loop paths around the target \cite{krovi2016}.
The performance of the algorithm is optimised again by tuning the
associated parameters.

\section{Has Evolution Exploited Advantages of Grover's Algorithm?}

Biological systems, especially at the molecular level, have two striking
features:\\
(a) Various biomolecules function according to their chemical and
structural properties.
Most of the time, they are not readily available to the living organism.
Rather a living organism eats food, breaks it down to its elementary
components by digestion, and then reassembles the components according
to specific prescriptions to obtain the required biomolecules.
In this metabolic process, the elementary components are randomly floating
around in the cellular environment, and the task of assembling them in a
specific order is that of unstructured database search.\\
(b) Living organisms are non-equilibrium systems, sustained against the
odds by clever manipulations of energy.
The tasks of efficient acquisition and transfer of energy have therefore
high priority.\\
Given that Grover's algorithm provides the optimal solution to both these
requirements, the simplicity, the robustness and the versatility of the
algorithm, and the persistent hunt of biological evolution to find ingenious
and efficient solutions to the problems at hand (i.e. survival of the fittest),
it would be a surprise if nature hadn't discovered Grover's algorithm, even
without a systematic analysis.
Indeed, the evidence described below highlights the manner in which Grover's
algorithm may have already become a key and an inalienable part of life.

Whenever a suggestion regarding the role of Grover's algorithm in a biological
process is made, immediate concerns are raised regarding how highly fragile
quantum dynamics can survive in the cellular environment with continuous
jostling of a large number of molecules.
The elaboration of the previous sections was to emphasise that the quantum
properties required for the execution of Grover's algorithm are rather
minimal, and can be replaced by appropriate classical wave analogues.
In particular, the following aspects are worth keeping in mind:\\
(1) In the biological context, time is highly precious while space is
fairly expendable, in sharp contrast to the conventional computational
complexity framework that treats time and space on an equal footing.
Biological systems can sense small differences in population growth rates,
and even an advantage of a fraction of a percent in time is sufficient for
one species to overwhelm another over many generations.
Spatial resources are frequently wasted, that too on purpose.
Just think of how many seeds a plant produces, when a single one can
ensure the continuity of its lineage.
(Note that such wastefulness also leads to competition and Darwinian
selection.)
Thus living organisms may be able to afford the classical version of
Grover's algorithm, with its enhanced stability compared to the quantum
version; the additional cost of spatial resources may remain tolerable
for small values of $N$.
That would beat the Boolean algorithm for the same task in the cost of
temporal resources, which is crucial for the biological tasks.
There can be even mixed scenarios, where fragile quantum steps are
stabilised by embedding them in a background classical evolution.\\
(2) Coherent superposition of wave modes, classical or quantum, is an
essential part of Grover's algorithm.
It must survive long enough for the algorithm to execute.
But the algorithm would still work if the superposition is merely apparent,
and not genuine.
That would happen if the cycling time between different states is short
compared to the time required to select the target state (e.g. the
appearance of spokes of a rapidly spinning wheel), which is possible when
molecular diffusion in cells is fast.
It is worth noting that the molecular coherence of biomolecules (e.g. a
polypeptide) can be delocalised over a region much larger than the molecular
size \cite{shayeghi2019}.\\
(3) Similarity of Grover's algorithm with localisation suggests that it is
possible for underlying quantum dynamics to produce macroscopic classical 
effects in interacting many-body systems.
Moreover, the dynamics is easier to protect from external disturbances
when the steps involved are not too many.

The unique signature of Grover's algorithm appears most strikingly in the
structure of genetic languages \cite{patel2001,patel2007}.
The languages of genes and proteins are universal for all living organisms,
they use specific building blocks (i.e. nucleotide bases and amino acids)
from the many similar ones available in cells, and the information they
carry is packed to nearly maximum density.
These properties indicate that these languages are essentially optimal
solutions for the tasks they carry out, and not just a frozen accident of
history.
During replication and translation, new DNA/RNA and polypeptide chains
are synthesised by sequentially assembling their building blocks in an
order specified by preexisting master templates.
The correct building block is identified by complementary base-pairing;
either it takes place or it does not.
Thus the problem solved is indeed unstructured database search using a
binary oracle provided by the master template.
The smallest three solutions of Grover's algorithm in this situation,
obtained from Eq.(\ref{querysoln}), are \cite{noninteger}:
\begin{equation}
Q=1 \rightarrow N=4 ~,~~
Q=2 \rightarrow N=10.5 ~,~~
Q=3 \rightarrow N=20.2 ~.
\end{equation}
These are in remarkable agreement with the identification of the four
nucleotide bases of DNA/RNA with a single base-pairing, the identification
of the twenty amino acids in polypeptide chains with a triplet code, and
the identification of ten amino acids in either of their two classes by a
doublet code \cite{patel2005}.
There is no other known scenario that explains these numbers as optimal
solutions to the actual information processing task accomplished \cite{optsol};
the best Boolean algorithm for the same task is binary search, which yields
$N=2^Q$.

The smallest instance, $N=4$, is an instructive example.
Its solution with Grover's optimal quantum algorithm requires one query
and two qubits.
When that is fragile and impractical, the Boolean search solves the
problem with two queries and two bits (i.e. an extra factor of two in
temporal resources), and the classical wave search solves it with one query
and four wave modes (i.e. an extra factor of two in spatial resources).
In biological systems, cheaper spatial resources than temporal ones
would obviously favour the latter solution.
Note that the square-root temporal gain offered by the classical wave
search over the Boolean search (i.e. $O(\sqrt{N})$ vs. $N$) is indeed
comparable to the exponential spatial cost involved (i.e. $N$ vs. $\log_2N$)
for small values of $N$.

If one imagines the development of a genetic information encoding system
when life originated, it would have been certainly sufficient and easier
to do the job using two nucleotide bases (one complementary pair) and
a Boolean algorithm.
Was it then the advantage of classical wave search that led nature to
complicate the encoding to the universal genetic languages observed today?
It is entirely plausible that some primitive organism discovered the
advantages of Grover's algorithm, quite likely by trial and error,
and built that at the core of life's information processing system.
Simple models that map the base-pairing to the reflection oracle and the
approach to equilibrium to an oscillatory process, incorporating the features
described in Section 4, can be constructed \cite{patel2001,patel2007}.
Nevertheless, for a realistic description that accurately identifies the
dynamical execution of the algorithm, we need experimental observations of
the intermediate steps of the genetic replication and translation processes.
That is not yet possible in sufficient detail, but the progress in technology
should take us there some day.
Alternatively, indirect checks that compare the efficiency of the natural
system with artificially constructed competitors (using different number
of letters in the genetic alphabet) would be easier to explore
\cite{patel2001a}.

Another biological phenomenon relevant to Grover's algorithm is the
process of energy transfer during photosynthesis, from the chlorophyll
pigment molecules that capture photons to the reaction centre where
glucose is synthesised.
This energy transfer is nearly dissipationless and takes place as coherent
wave motion of an exciton in a network of pigment molecules \cite{fleming2009}.
A classical strategy of hopping in a funnel-shaped energy landscape
cannot explain this behaviour.
But it can be understood as amplitude amplification in a spatial search
algorithm, with the reaction centre acting as a defect that induces
localisation, and the process being terminated when the accumulated
energy crosses the threshold for ionising water.
Models with effective Hamiltonians for the pigment network have been
constructed, but details of the process in presence of the existing
environmental noise still remain to be properly understood.
A rough analogy would be how a crack in an object, say a child's toy,
opens up, when it is shaken in a suitable manner---the crack reflects the
wave motion causing energy to build up there, and then nonlinear material
dynamics cascades the energy down to the atomic scale where bonds are broken.

Telltale signatures of vibronic modes (i.e. coupled vibrational and
electronic degrees of freedom) have also been found in enzyme catalysis,
olfaction and magnetoreception by birds \cite{huelga2013}.
A standard test for significant contribution from the vibrational degrees of
freedom in such phenomena is the kinetic isotope effect \cite{westheimer1961},
where isotopic substitution alters the vibrational properties of molecules
with negligible effect on the electronic structure.
How the involvement of vibrational modes and wave dynamics can help these
processes is a topic of active research.
Needless to say, a better understanding of such processes discovered by
natural evolution, combined with features of Grover's algorithm, would
allow us to design new types of catalysts and sensors.
Moreover, when the turn comes to develop quantum memories, to go along
with the quantum devices being developed, organising them with quaternary
addresses (instead of binary ones) will be an attractive proposition worth
considering.

\section*{Acknowledgments}
It is a pleasure to acknowledge many stimulating interactions with
Lov Grover over the years.


\begin{thebibliography}{9}
\bibitem{evolop}{Unitary evolution operators are interpreted as
        $U=e^{-iH\tau}$, and $R=1-2P=e^{i\pi P}$.}
\bibitem{noninteger}{In reality, both $Q$ and $N$ are integers.
        When a solution to Eq.(\ref{querysoln}) has a non-integer value
        for $N$, it means that the result produced by the algorithm will
        have a small error.}
\bibitem{optsol}{There is no other explanation for the four-letter alphabet
        of DNA/RNA. Assuming this four-letter alphabet, some scenarios to
        explain the twenty-letter amino acid alphabet have been constructed,
        but they need inputs beyond just optimisation of information
        processing.}
\end{thebibliography}
\end{document}